\documentclass[runningheads]{llncs}
\usepackage[T1]{fontenc}
\usepackage{graphicx}
\usepackage{comment}
\usepackage{amsmath}
\usepackage{tabularray}
\usepackage{multicol}
\usepackage{multirow}
\usepackage{gensymb}
\usepackage{lipsum}
\usepackage{pythontex} 
\usepackage{nopageno}
\usepackage{graphicx}
\usepackage{subcaption}
\usepackage{graphicx}
\newsavebox\CBox
\def\textBF#1{\sbox\CBox{#1}\resizebox{\wd\CBox}{\ht\CBox}{\textbf{#1}}}

\begin{document}
    \title{GPC: Generative and General Pathology Image Classifier}
    \author{Anh Tien Nguyen \and  Jin Tae Kwak}
    \authorrunning{A.T. Nguyen \&  J.T. Kwak}
    \institute{School of Electrical Engineering, Korea University, Seoul 02841, Korea
    \email{\{ngtienanh,jkwak\}@korea.ac.kr}
    }
    
    \maketitle              
    
    \begin{abstract}
        Deep learning has been increasingly incorporated into various computational pathology applications to improve its efficiency, accuracy, and robustness. Although successful, most previous approaches for image classification have crucial drawbacks. There exist numerous tasks in pathology, but one needs to build a model per task, i.e., a task-specific model, thereby increasing the number of models, training resources, and cost. Moreover, transferring arbitrary task-specific model to another task is still a challenging problem. Herein, we propose a task-agnostic generative and general pathology image classifier, so called GPC, that aims at learning from diverse kinds of pathology images and conducting numerous classification tasks in a unified model. GPC, equipped with a convolutional neural network and a Transformer-based language model, maps pathology images into a high-dimensional feature space and generates pertinent class labels as texts via the image-to-text classification mechanism. We evaluate GPC on six datasets for four different pathology image classification tasks. Experimental results show that GPC holds considerable potential for developing an effective and efficient universal model for pathology image analysis.
    \end{abstract}

    \section{Introduction}
        In computational pathology, pathological image classification has been extensively studied~\cite{cui2021artificial}. There exist various kinds of image classification tasks such as cancer detection, cancer grading, and tissue typing \cite{multi_scale,joint,gastric_trinh,impash}. These tasks are essential in pathology since they are closely related to decision-making in patient care and treatment. In clinics, these routine works suffer from inefficiency, inaccuracy, and variations, particularly with the increase in the workload per pathologist~\cite{pathologist_trend}. In recent years, machine learning and artificial intelligence techniques have been increasingly applied to pathology image analysis and shown to be effective in such tasks. Many of such methods adopt convolutional neural networks (CNNs) \cite{multi_scale,joint,gastric_trinh} and, more recently, Transformer-based models have been often employed for differing tasks \cite{fu2022stohisnet,wang2021transpath}. Although both CNN and Transformer-based models have shown to be promising in analyzing pathology images, there is one drawback with these approaches. There exist numerous tasks in pathology that are closely related to each other; for instance, cancer grading in different types of organs such as the prostate, colon, gastric, and breast. With the current approaches, one needs to develop a separate model per task, which is challenging to transfer a pre-existing model to other related tasks. To tackle such a problem, a unified or general model that can simultaneously process different types of images and conduct multiple tasks on them is needed.

        Therefore, we introduce a task-agnostic \textBF{G}enerative and general \textBF{P}athology image \textBF{C}lassifier (GPC) that can process and learn from arbitrary datasets of pathology images and perform multiple image classification tasks in a generative image-to-text fashion. To the best of our knowledge, this is the first attempt to build a generative and general image-to-text classifier for pathology images. GPC exploits the recent developments of CNNs and Transformer-based language models. Given a pathology image $x$, it produces a high-level feature representation by CNN and generates the pertinent class label as a text by a language model, which is built based upon Transformers. Since GPC utilizes the language model, it can handle different types of images and tasks at the same time. To evaluate the proposed GPC, we integrate four separate pathology classification tasks: colorectal cancer grading, prostate cancer grading, gastric cancer grading, and colorectal tissue typing. By employing pathology images from different types of organs and tasks, we aim to improve the utility of the existing pathology images and task-specific ground truth labels, to learn organ- and task-agnostic representations of pathology images, and to strengthen the predictive power of the pathology image classifier. The experimental results demonstrate that GPC can facilitate a unified and general image classification for pathology images.


    \section{Methodology}
        \subsection{Problem formulation}
            
            Suppose that we are given $M$ datasets $\{D_1, D_2, ..., D_M\}$, $D_i=\{(x^{i,k}, c^{i,k})\}_{k=1}^{N}$, where $x^{i,k}$ and $c^{i,k}$ denote the $k$-th pathology image and its ground truth in the $i$-th dataset, respectively. Since $c^{i,k}$ is a text label such as \textit{benign} and \textit{poorly-differentiated cancer}, we split and pad it into a sequence of tokens $t^{i,k}$ using a tokenizer of a language model $\mathcal{L}$. As a result, each dataset is modeled as $D_i=\{(x^{i,k}, (t^{i,k}_1, t^{i,k}_2, ..., t^{i,k}_T)\}_{k=1}^{N}$, where $T$ is the maximum length of the token sequence of all text labels.

            Each pathology image $x^{i,k}$ undergoes a feature extractor $\mathcal{F}$ and a projector $\mathcal{P}$ to produce a feature embedding $f^{i,k}$ as follows:
                \begin{equation}
                    f^{i,k} = \mathcal{P}(\mathcal{F}(x^{i,k}))
                \end{equation}
            Afterward, the projected embedding $f^{i,k}$ is used as a condition for the language model $\mathcal{L}$ to generate a text label autoregressively, i.e., predicting the next token given the previously generated tokens. 
            Specifically, at each step, the next token is drawn from the probability distribution over the vocabulary that is based on the concatenation of $f^{i,k}$ and the embeddings of the previous tokens. We employ the greedy approach in which the token with the highest probability is selected as the output:
                \begin{equation}
                    \tilde{t}^{i,k}_m =\underset{\hat{t}^{i,k}_m}{\arg\max} \: p(\hat{t}^{i,k}_m | f^{i,k}, \tilde{t}_{1}^{i,k},\tilde{t}_{2}^{i,k},...,\tilde{t}_{m-1}^{i,k})
                \end{equation}
            where $\tilde{t}^{i,k}_m$ refers to the $m$-th predicted token for the $k$-th pathology image in the $i$-th dataset.
            As a result, the objective of our study can be formulated as following:
                \begin{equation}
                    \theta = \underset{\hat{\theta}}{\arg\max} \sum_{i=1}^{M} \sum_{k=1}^{N}\sum_{m=1}^{T}  \log p_{\hat{\theta}}(\tilde{t}_{m}^{i,k}|f^{i,k},\tilde{t}_{1}^{i,k},\tilde{t}_{2}^{i,k}...,\tilde{t}_{m-1}^{i,k})
                \end{equation}
            where $\theta$ represents the learnable parameters of GPC.
    
        \subsection{Network architecture}
\begin{figure}
                \includegraphics[width=\textwidth]{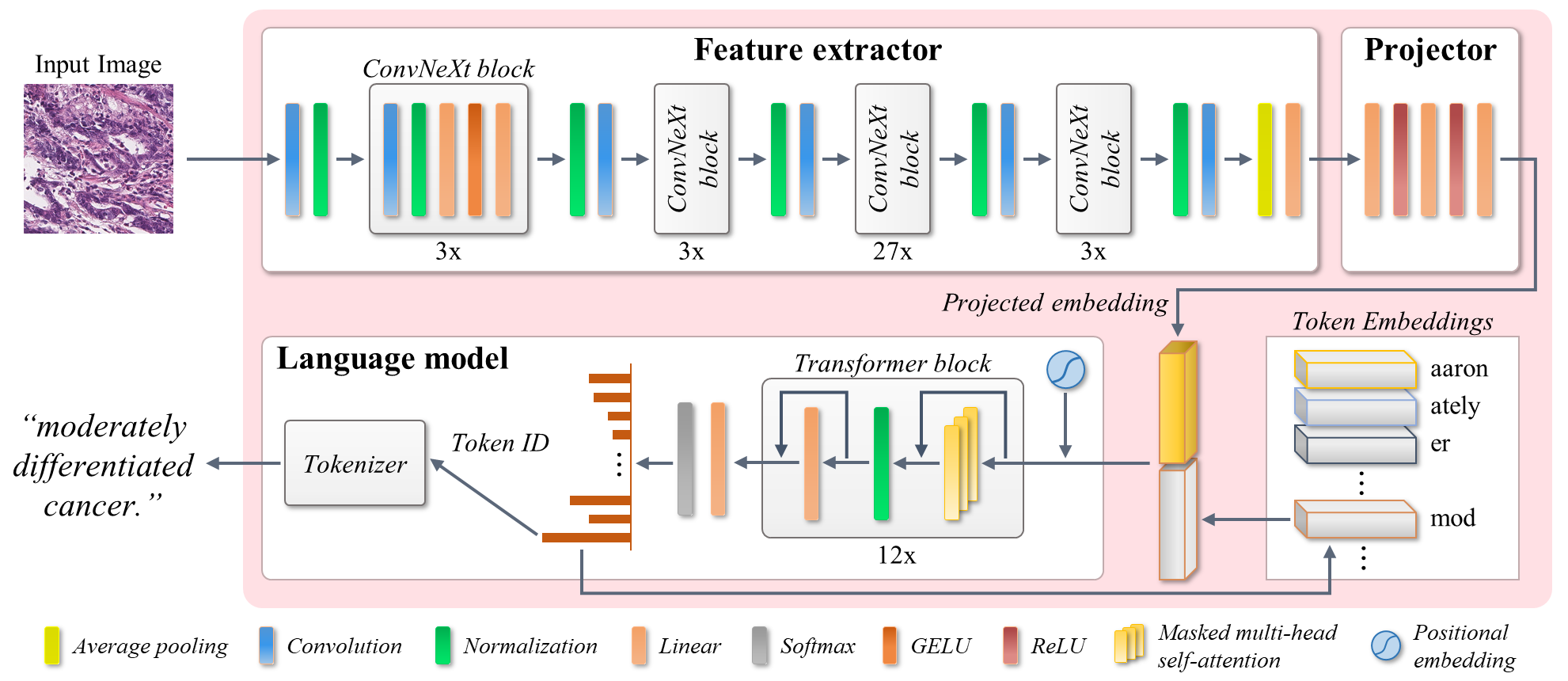}
                \caption{Overview of GPC.} \label{model}
            \end{figure}

        The overview of GPC architecture is illustrated in Fig.~\ref{model}. GPC consists of three primary components: 1) a feature extractor $\mathcal{F}$, 2) a projector $\mathcal{P}$, and 3) a language model $\mathcal{L}$. 
            
            \subsubsection{Feature extractor} Feature extractor $\mathcal{F}$ is built based upon a CNN to extract high-level representations from input images. We utilize ConvNeXt due to its excellent performance in image classification, comparable to recent Transformer-based approaches while retaining the effectiveness of a simple architecture of CNNs. It employs ResNet-50~\cite{resnet} as a backbone and transforms the architecture following the design strategies of Transformers by investigating and adopting several design techniques step-by-step, including macro and micro design, ResNeXt~\cite{resnext}, inverted bottleneck, and large kernel size.

            \subsubsection{Projector} Projector $\mathcal{P}$ maps the output features from the space of $\mathcal{F}$ to that of a language model $\mathcal{L}$. It simply utilizes a multilayer perceptron with three fully-connected layers. $\mathcal{P}$ bridges the gap between the image feature domain and the text feature domain in a way that the image feature embeddings of $\mathcal{F}$ guided and adjusted to align with the feature space of $\mathcal{L}$.
    
            \subsubsection{Language model} Language model $\mathcal{L}$ generates the correct pathological text labels for the projected embeddings obtained from $\mathcal{P}$. We select Open Pre-trained Transformer language models (OPT)~\cite{opt} as $\mathcal{L}$. Since OPT is a decoder-only pre-trained Transformer-based language model, it easily applies to image-to-text generation tasks. We chose the base version of OPT among several variants, including a stack of 12 Transformer layers with 12-head attention layers due to computational complexity and cost.

    \renewcommand{\arraystretch}{1.3}

    \section{Experiments}
    
        \subsection{Datasets} 
            We investigate six datasets of four pathology image classification tasks: 1) colorectal cancer grading, 2) prostate cancer grading, 3) gastric cancer grading, and 4) colorectal tissue typing. The details of the datasets are shown in Table~\ref{dataset_tab}. 
            
            \subsubsection{Colorectal cancer grading}: 
            Two public datasets (Colon-1 and Colon-2) are collected from~\cite{joint}. Colon-1 and Colon-2 include 9,857 patch images and 110,170 patch images, respectively. Each image is assigned a class label, including \textit{benign}, \textit{well differentiated cancer}, \textit{moderately differentiated cancer}, and \textit{poorly differentiated cancer}. Colon-1 is split into a training, validation, and test set. Colon-2 is utilized as an independent test set.

            \subsubsection{Prostate cancer grading}: 
            We utilize two public prostate datasets (Prostate-1 and Prostate-2). Prostate-1 was obtained from the Harvard dataverse (https:// dataverse.harvard.edu). Prostate-2 was acquired from Gleason2019 challenge (https://gleason2019.grand-challenge.org). Both are annotated with four class labels: \textit{benign}, \textit{grade 3 cancer}, \textit{grade 4 cancer}, and \textit{grade 5 cancer}. Prostate-1 contains 22,022 patch images that are split into a training, validation, and test set. Prostate-2 has 17,066 patches that are used as an independent test set.

            \subsubsection{Gastric cancer grading}: 
            A single gastric cancer dataset (Gastric) was obtained from a local hospital. It includes 265,066 patch images with four class labels, including \textit{benign}, \textit{tubular well differentiated cancer}, \textit{tubular moderately differentiated cancer}, and \textit{tubular poorly differentiated cancer}. The entire dataset is split into a training, validation, and testing set.
         
            \subsubsection{Colorectal tissue typing}:
            A publicly available dataset (K19) was attained from~\cite{k19}. K19 includes 100,000 images that are categorized into \textit{adipose}, \textit{background}, \textit{debris}, \textit{lymphocyte}, \textit{normal}, \textit{stroma}, \textit{epithelium}, \textit{muscle}, and \textit{mucus}. K19 is divided into a training, validation, and testing set.

            \begin{center}
                \begin{table}[!t]
                \caption{Details of datasets. TR, VAL, and TS denote training, validation, and test sets, respectively.}\label{dataset_tab}
                \resizebox{\textwidth}{!}{
                \begin{tabular} {ccccc}
                    \hline
                    Task & Dataset & Mag. & Patch Size & \# Patches \\ \hline
                    \multirow{2}{*}{Colorectal cancer grading} & Colon-1 & 20x & 512 x 512 & TR (7,027), VAL (1,242), TS-1 (1,588) \\ 
                           
                        & Colon-2 & 20x & 512 x 512 & TS-2 (110,170) \\

                    \multirow{2}{*}{Prostate cancer grading} & Prostate-1 & 40x & 750 x 750 & TR (15,303), VAL (2,482), TS-1 (4,237) \\
                    & Prostate-2 & 40x & 690 x 690 & TS-2 (17,066) \\

                    Gastric cancer grading & Gastric & 40x & 512 x 512 & TR (233,898), VAL (15,381), TS (15,787) \\ 
                    Colorectal tissue typing & K19 & 20x & 224 x 224 & TR (70,000), VAL (15,000), TS (15,000) \\ 
                    \hline
                    \end{tabular}}
                \end{table}
            \end{center}

        \subsection{Comparative models}
            We compare three other types of models with GPC. The models include 1) three CNN models: ConvNeXt-L~\cite{convnext}, EfficientNetV2-S~\cite{efficientnetv2}, and ResNet50~\cite{resnet}, 2) three Transformer models: MaxViT~\cite{maxvit}, SwinV2-B~\cite{swin2}, and ViT-B~\cite{vit}, and 3) two generative models: CLIP~\cite{clip} and GIT-B~\cite{git}. GIT-B is an end-to-end Transformer-based model for image captioning that are similar to our approach. Regarding CLIP, we only obtain the pre-trained vision branch of CLIP-ViT-L-14 as an image extractor and integrate OPT-125M as a text decoder.

            \begin{figure*}[t!]
                \includegraphics[width=\textwidth]{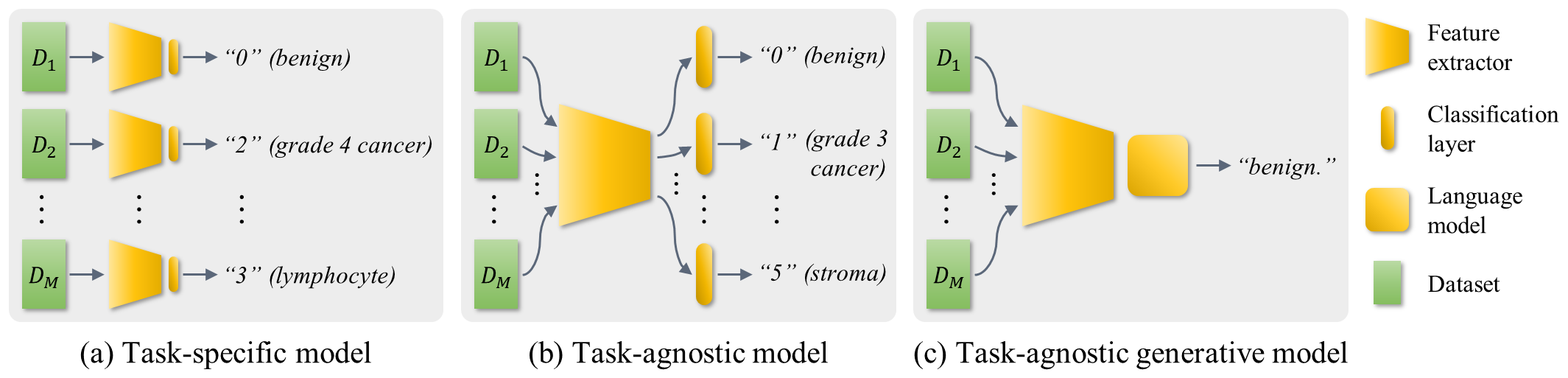}
                \caption{Three types of classification approaches. (a) Task-specific model, (b) Task-agnostic model, and (c) Task-agnostic generative model.} \label{compare}
            \end{figure*}
    
        \subsection{Experimental design}
            We conduct three settings to assess GPC and competitors (Fig.~\ref{compare}): 1) Task-specific classification ($E_{TS}$): A model, equipped with a feature extractor and a classifier head, is trained on a training set and tested on the test set(s) per classification task, 2) Task-agnostic classification ($E_{TA}$): A model contains a feature extractor and four classifier heads, i.e., one classifier head per task. It is trained on all training sets from four classification tasks and assessed on each test set per task using the corresponding classifier, and 3) Task-agnostic generative classification ($E_{TAG}$): A model includes a feature extractor and a generative classifier. It is trained on all training sets from four classification tasks and evaluated on the entire test sets. Three CNNs and three Transformer models are utilized in $E_{TS}$ and $E_{TA}$. In $E_{TA}$, the training loss is only aggregated by the output of the corresponding classification layer. Two generative models and GPC are used in $E_{TAG}$.

        \subsection{Training details}
            During training, five data augmentation techniques are implemented: random horizontal flip, affine transformation, image blurring, random additive Gaussian, and random color change. The first three techniques are applied to every patch, while the others have a 50\% chance of being applied. After applying data augmentation techniques, all image patches are resized to dimensions of 512 × 512 pixels. AdamW~\cite{adamw} is adopted to optimize the learnable parameters with a learning rate of $10^{-5}$ controlled by a cosine annealing warm restarts scheduler~\cite{scheduler} during 60 epochs. 

        \subsection{Metrics}
            To evaluate the performance of GPC and other competitors, we measure several evaluation metrics. For cancer grading, we use accuracy ($Acc$), cancer grading accuracy ($Acc_g$), macro-averaged F1 ($F1$), and quadratic weighted kappa ($k_w$)~\cite{kappa}. Regarding tissue typing, we calculate $Acc$, $F1$, macro-averaged precision ($Pre$), and macro-averaged recall ($Re$). 

        \begin{table}[!t]
            \caption{Results of colorectal cancer grading and tissue typing.}\label{colon_result}
            \resizebox{\textwidth}{!}{
            
            \begin{tabular} {lccccccccccccccc}
                \hline
                \multirow{2}{*}{Method} & \multirow{2}{*}{Type} & \multicolumn{4}{c}{Colon-1}    &  & \multicolumn{4}{c}{Colon-2} &  & \multicolumn{4}{c}{K19} \\ \cline{3-6}  \cline{8-11} \cline{13-16}
                 &    & $Acc$ (\%) &  $Acc_g$ (\%) & $F1$     & $K_w$  &  & $Acc$ (\%) &  $Acc_g$ (\%) & $F1$   & $K_w$  &  & $Acc$ (\%) & $Pre$   & $Re$ & $F1$ \\ 

                 \hline

                ConvNeXt-L & \multirow{6}{*}{$E_{TS}$} & 87.7  & 82.8 & 0.832   & 0.940 &  & 78.1	& 71.9 & 0.731	& \textBF{0.908} & & \textBF{99.6} & \textBF{0.996}   & \textBF{0.996} &  0.994     \\
                EfficientNetV2-S & & 85.9  & 80.9 & 0.819   & 0.914 &  & 76.9 & 68.4 & 0.708 & 0.701 & & 98.0 & 0.973   & 0.968 &  0.985  \\
                ResNet50  &       & 86.8  & 82.9 & 0.838   & 0.806 &  & 79.5 & 68.2 & \textBF{0.733} & 0.688 & & 98.7 & 0.988   & 0.988 &  0.987  \\
                MaxViT     &       & 87.9  & \textBF{84.0} & 0.838   & 0.805 &  & 76.3 & 72.8 & 0.723 & 0.895 & & 98.3 & 0.988   & 0.991 &  0.988  \\
                SwinV2-B  &       & 88.0  & 82.7 & 0.829   & 0.839 &  & 77.9 & 73.7 & 0.729 & 0.885 & & 99.4 & \textBF{0.996}  & 0.993 & 0.991  \\
                ViT-B      &       & 87.5 & 82.0 &  0.838   & 0.838 &  & \textBF{79.8} & 72.8 & 0.728 & 0.899 & &  98.2 & 0.989   & \textBF{0.996} & 0.988 \\ \hline

                ConvNeXt-L & \multirow{6}{*}{$E_{TA}$} & 85.9 & 80.4 &  0.823   & 0.933 &  & 74.4 & 66.5  & 0.698	& 0.868 & & 98.8 & 0.986   & 0.991 & 0.988 \\
                EfficientNetV2-S & & 83.2 & 79.2 &  0.793   & 0.882 &  & 72.5 & 63.4 & 0.670    & 0.722 & & 98.5 & 0.982   & 0.982 & 0.974  \\
                ResNet50  &      & 84.1 & 81.0 &  0.807   & 0.824 &  & 70.1 & 61.9 & 0.622    & 0.671 & & 97.7 & 0.984  & 0.983& 0.986  \\
                MaxViT     &      & 86.8 & 82.6 &  0.809   & 0.813 &  & 71.3 & 68.8 & 0.720  & 0.888 &  & 98.3 & 0.985   & 0.991 & 0.974  \\
                SwinV2-B  &      & 86.5 & 81.2 &  0.822   & 0.933 &  & 70.4 & 69.0 & 0.671  & 0.842 & & 98.4 & 0.985   & 0.980 & \textBF{0.996}  \\
                ViT-B      &      & 86.0  & 79.7 & 0.812   & 0.831 &  & 72.1 & 67.1 & 0.701  & 0.833 & & 98.1 & 0.985   & 0.989 & 0.988  \\
                
                \hline
                GIT       & \multirow{3}{*}{$E_{TAG}$} & 85.3  & 79.7 & 0.811   & 0.924 &  & 67.9 & 58.6 & 0.596 & 0.839 & & 98.9 & 0.989   & 0.988 & 0.990 \\
                CLIP+OPT & & 82.5  & 75.6 & 0.795   & 0.914 &  & 72.7 & 67.4 & 0.653 & 0.791 & & 99.0 & 0.989   & 0.992 & 0.985  \\ 
                GPC (ours)  &   & \textBF{88.4}	& 83.8 & \textBF{0.848}   & \textBF{0.944} &  & 79.0 & \textBF{74.0} & 0.722 & 0.898 & & 99.4 & 0.995  & 0.995 & \textBF{0.996} \\ \hline
            \end{tabular}}
        \end{table}

        \begin{table}[!t]
            \caption{Results of prostate and gastric cancer grading.}\label{prostate_result}
            \resizebox{\textwidth}{!}{
            \begin{tabular}{lccccccccccccccc}
                \hline
                \multirow{2}{*}{Method} & \multirow{2}{*}{Type} & \multicolumn{4}{c}{Prostate-1}    &  & \multicolumn{4}{c}{Prostate-2} &  & \multicolumn{4}{c}{Gastric} \\ \cline{3-6}  \cline{8-11} \cline{13-16}
                &  & $Acc$ (\%) & $Acc_g$ (\%)  &  $F1$   & $K_w$  &  & $Acc$ (\%) & $Acc_g$ (\%) & $F1$  & $K_w$ &  & $Acc$ (\%) & $Acc_g$ (\%) & $F1$   & $K_w$  \\ 
                \hline

                ConvNeXt-L  & \multirow{6}{*}{$E_{TS}$}  & 70.6 & 70.1 &  0.630   & 0.597 &  & 77.8 & 78.2 & 0.639 & 0.696 & & 83.8 & 68.1 &  0.760   & 0.925 \\ 
                EfficientNetV2-S &     & 69.7  & 66.4 & 0.582  & 0.504 &  & 74.3 & 77.3 & 0.599 & 0.633 & &81.3 & 68.1 &  0.712   & 0.890\\
                ResNet50   &  & 70.9 & 67.5 &  0.643   & 0.512 &  & \textBF{77.3} & 78.7 & 0.608 & 0.619 & &82.2 & 66.9 &   0.707   & 0.901 \\
                MaxViT      &  & 71.6 & 70.2 &  \textBF{0.652}  & \textBF{0.649} &  & 75.9 & 76.7 & 0.605 & 0.678  & &83.2 & 68.5 &  0.758   & 0.926\\
                Swin-V2-B   &  & \textBF{71.9} & 72.0 &  0.637   & 0.639 &  & 73.9 & 75.1 & 0.623 & 0.669  & &83.9 & 68.5 & 0.771   & \textBF{0.935}\\
                ViT-B       &  & \textBF{71.9} & \textBF{72.2} &  0.641   & 0.643 &  & 75.4 & 75.9  & 0.608 & 0.690 & &\textBF{84.4} & 69.2 &  \textBF{0.774}   & 0.930 \\ \hline

                ConvNeXt-L  & \multirow{6}{*}{$E_{TA}$}  & 68.5 & 69.7 &  0.576   & 0.578 &    & 73.3 & 76.3 & 0.562 & 0.616 & & 83.0 & 67.2 &  0.757   & 0.930 \\
                EfficientNetV2-S  &    & 65.2 & 62.1 &  0.522   & 0.511 &    & 71.9 & 73.1 & 0.512 & 0.589 & & 80.5 & 63.5 & 0.701   & 0.832 \\
                ResNet50    &  & 69.2 & 68.1 &  0.582   & 0.539 &    & 73.6 & 77.3 & 0.599 & 0.601 & & 82.9 & 64.0 & 0.713   & 0.890 \\
                MaxViT       &  & 67.2  & 69.2 &  0.606   & 0.562 &    & 69.2 & 70.9 & 0.525 & 0.631 & & 83.6 & 65.9 &  0.749   & 0.931 \\
                Swin-V2-B    &  & 65.5  & 66.9 & 0.531   & 0.542 &    & 65.8 & 69.2 & 0.487 & 0.553 & & 81.7 &  65.0 & 0.739   & 0.923\\
                ViT-B        &  & 67.2  & 66.4 & 0.544   & 0.579 &    & 68.8 & 72.8 & 0.598 & 0.629 & & 81.7 &  64.2 & 0.710   & 0.909\\ \hline
                
                GIT          &  \multirow{3}{*}{$E_{TAG}$}   & 65.9 & 67.2 &  0.538   & 0.476 &  & 68.3 & 71.7 & 0.467 & 0.616 & &80.7 & 63.7 &  0.727   & 0.867 \\
                CLIP+OPT     &  & 62.0  & 63.3 &  0.598   & 0.587 &  & 63.4 & 62.2 & 0.521 & 0.575 & & 81.6 & 63.4 &  0.726   & 0.912\\ 
                GPC (ours)   &  & 70.4 & 71.9  &  0.628   & 0.612 &  & 76.9 & \textBF{79.0} & \textBF{0.641} & \textBF{0.700} & & 83.7 & \textBF{69.3} & 0.768   & 0.925 \\ \hline
            \end{tabular}}
        \end{table}

    \section{Results and Discussion}

        We conduct four classification tasks with six pathology image datasets using GPC and other competitors. The competitors include three CNN models and three Transformer models with two different experimental settings ($E_{TS}$ and $E_{TA}$) and two generative models, i.e., GPC is compared with 14 different models. Table~\ref{colon_result} and~\ref{prostate_result} demonstrate the experimental results of four classification tasks. For colorectal cancer grading, GPC outperforms all competitors on Colon-1. In terms of Colon-2, it obtains the best $Acc_g$ and ranks top-3 for $Acc$ and $Kw$. For other metrics, there is no consensus. ViT-B, ResNet50, and ConvNeXt-L in $E_{TS}$ achieved the best $Acc$, $F1$, and $Kw$, respectively. In prostate cancer grading, though GPC is sub-optimal for Prostate-1 (top-4 in $Acc_g$ and $Kw$ and top-6 in $Acc$ and $F1$), it outperforms other competitors on three out of four evaluation metrics on Prostate-2. As for gastric cancer grading, GPC is ranked first in $Acc_g$, third in $F1$, and fourth in $Acc$. ViT in $E_{TS}$ obtains the best $Acc$ and $F1$. In colorectal tissue typing, GPC achieves the best $F1$ and second best $Acc$, and is only short by 0.001 for $Pre$ and $Re$.

        In a head-to-head comparison between $E_{TS}$ and $E_{TA}$, models in $E_{TS}$ generally outperform those in $E_{TA}$. Two generative models (GIT and CLIP+OPT) are inferior to most of the CNN and Transformer models in both $E_{TS}$ and $E_{TA}$. It demonstrates the difficulty of fine-tuning a universal model for different classification tasks in pathology images. In the conventional deep learning approaches, a task-specific model is better suited for developing a model per task, which substantially increases the number of models and resources, limiting the scalability of the methods. GPC is not the best model for all classification tasks and datasets. Nonetheless, it achieved the best performance on two datasets (Colon-1 and Prostate-2) and was comparable to the best-performing models on four other datasets. It is also worth noting that there was no consensus on the best performing model for those four datasets. Hence, overall, GPC is the best model across the four classification tasks and six datasets.

    \begin{figure*}[t!]
        \includegraphics[width=\textwidth]{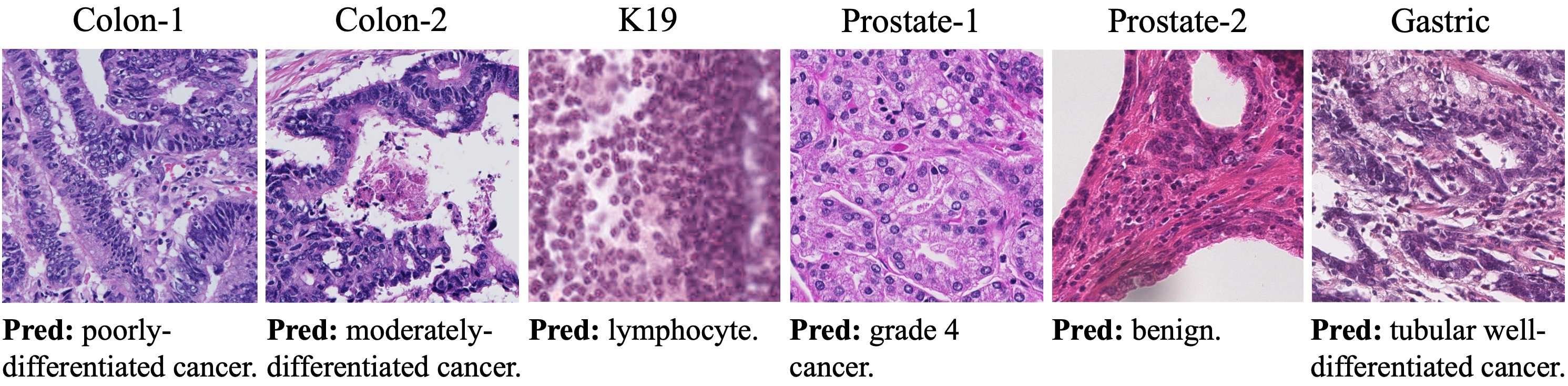}
        \caption{Examples of correct predictions by GPC. Pred denotes prediction.} \label{fig_true}
    \end{figure*}

    \begin{figure*}[t!]
        \includegraphics[width=\textwidth]{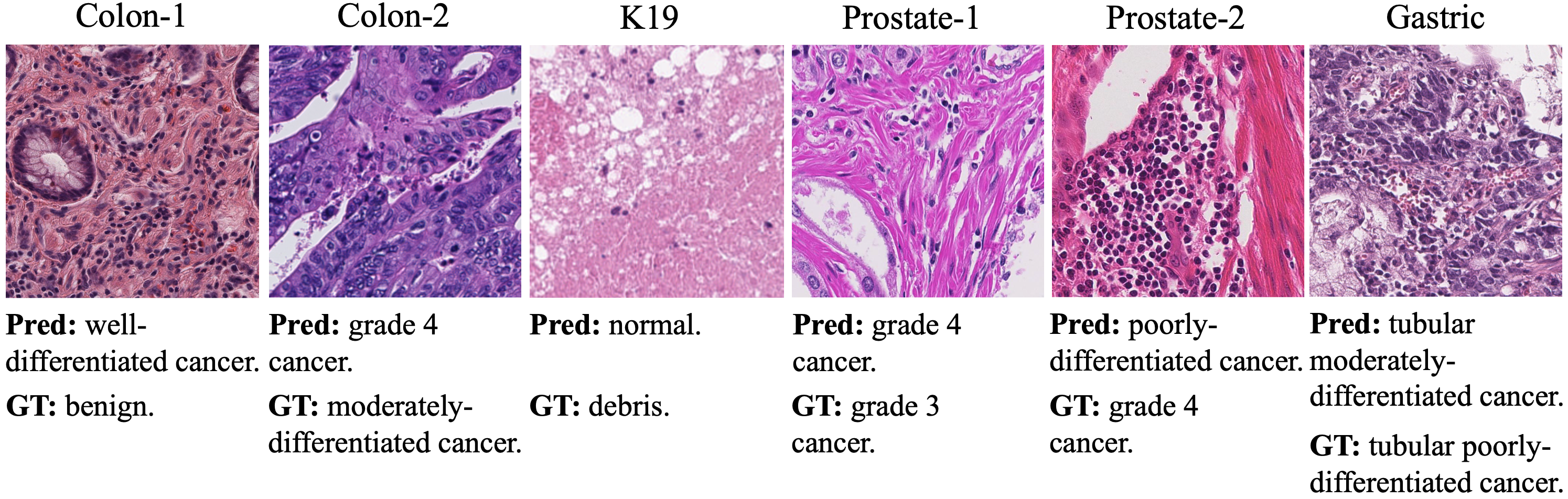}
        \caption{Examples of incorrect predictions by GPC. Pred and GT denote a prediction and a ground truth, respectively.} \label{fig_false}
    \end{figure*}

        Fig.~\ref{fig_true} depicts the exemplary samples correctly classified by GPC. Without the information of which organ each image was obtained from, GPC is able to predict and generate the correct class labels for differing cancer and tissue types. 
        Fig.~\ref {fig_false} shows the incorrect classification examples by GPC. For most such samples, GPC predicts in-domain labels, and cancer samples are classified as cancer, not benign, but of differing grades. 

        Table~\ref{model_compare} demonstrates the model complexity of GPC and other competing models in terms of floating point operations per second (FLOPS), number of parameters (millions), training time (milliseconds per image), and inference time (milliseconds per image) for the four classification tasks. 
        CNN and Transformer models, in general, contain a smaller number of parameters and FLOPS and a shorter amount of time for training and inference. Since GPC and other generative models adopt a visual encoder and a text decoder, they require a substantial amount of resource for training in particular; however, the inference time of GPC and other generative models is still $<$1 second per image.

    \begin{table}[!t]
    \caption{Model complexity of GPC and other models.}\label{model_compare}
    \setlength{\tabcolsep}{5pt} 
    \begin{center}
    \begin{tabular}{lccccc}
    \hline
    \multirow{2}{*}{Model} & \multirow{2}{*}{Type}  & FLOPS & Parameters & Training & Inference \\ 
     & & (B) & (M) & (ms/image) & (ms/image) \\
    \hline
    ConvNeXt-L      & \multirow{6}{*}{$E_{TA}$}   & 179.6  & 196.3   & 924.5    & 786.7       \\
    EfficientNetV2-S &  & 15.0     & 20.5    & 204.5    & 89.9           \\
    ResNet50         &  & 21.5     & 24.2    & 263.1    & 164.8         \\
    MaxViT           &  & 27.9     & 31.9    & 370.2    & 129.1           \\
    Swin-V2-B        &  & 53.4     & 87.6    & 477.0    & 391.0          \\
    ViT-B            &  & 17.6     & 86.6    & 639.3    & 220.2          \\ \hline
    GIT          & \multirow{3}{*}{$E_{TAG}$}    & 211.9    & 129.2   & 856.3    & 592.3           \\
    CLIP+OPT     &    & 263.8    & 427.7   & 1117.2   & 923.4           \\
    GPC (ours)   &    & 234.2    & 332.3   & 1088.6   & 870.5          \\ \hline
    \end{tabular}
    \end{center}
    \end{table}

    \section{Conclusions}
        In this study, we propose a generative and general pathology image classifier called GPC, which simultaneously learns and conducts multiple classification tasks with a single classification model. The experimental results demonstrate that the generative models, i.e., (pre-trained) language models, hold great potential for pathology image analysis, paving the way for developing a universal model for computational pathology. The future study will entail further development of generative models and extended validation on differing organs and tasks.

    \subsubsection{Acknowledgements} 
        This work was supported by the grant of the National Research Foundation of Korea (NRF) (No. 2021R1A2C2014557).
    
    \bibliographystyle{splncs04}
    \bibliography{references}
\end{document}